\documentclass[useAMS]{mn2e}

 \usepackage[usenames,dvipsnames]{color}

\usepackage{epstopdf}
\usepackage{graphicx}
\def \mdot   {{\hbox{$\skew3\dot M$}}}

\title[Structure in PN fast winds]{PN fast winds: Temporal 
structure and stellar rotation}
\author[R.K. Prinja et al.]{R. K. Prinja$^{1}$\thanks{E-mail: 
rkp@star.ucl.ac.uk (RKP)},
D.L. Massa$^{2}$, M. A. Urbaneja$^{3}$, R.-P. Kudritzki$^{3}$\\
$^{1}$Dept. of Physics {\&} Astronomy, University College London, Gower Street, London WC1E 6BT \\
$^{2}$Space Telescope Science Institute, 3700 San Martin Drive, Baltimore, 
MD 21218, USA \\
$^{2}$Institute for Astronomy, University of Hawaii, 2680 Woodlawn Drive, 
Honolulu, HI 96822, USA \\
}
\begin{document}

\date{Accepted 2009. Received 2009; in original form 2009}

\pagerange{\pageref{firstpage}--\pageref{lastpage}} \pubyear{2007}

\maketitle

\label{firstpage}

\begin{abstract}
To diagnose the time-variable structure in the fast winds of
PN central stars (CSPN), we present an analysis of
P~Cygni line profiles in {\it FUSE} satellite far-UV spectroscopic
data.
Archival spectra are retrieved to form time-series datasets for the
H-rich CSPN NGC~6826, IC~418, IC~2149, IC~4593 and NGC~6543.
Despite limitations due to the fragmented sampling of the time-series,
we demonstrate that in all 5 CSPN the UV resonance lines are
variable primarily due to the occurrence of blueward migrating
discrete absorption components (DACs).
Empirical (SEI) line-synthesis modelling is used to determine
the range of fluctuations in radial optical depth, which are
assigned to the temporal changes in large-scale wind structures.
We argue that DACs are common in CSPN winds, and their empirical
properties are akin to those of similar structures seen in the
absorption troughs of massive OB stars.
Constraints
on PN central star rotation velocities are derived from
Fast-Fourier Transform analysis
of photospheric lines for our target stars.
Favouring the causal role of co-rotating
interaction regions,
we explore connections between normalised DAC
accelerations and
rotation rates of PN central stars and O stars. 
The comparative properties suggest that the same physical mechanism
is acting to generate large-scale structure
in the line-driven winds in the two different settings.

\end{abstract}


\section{Introduction and background}
The central stars of planetary nebulae (CSPN) mark a critical stage in
the evolution of low to intermediate mass stars, and thus the majority
of stars in our Galaxy. Of significant interest are the fast winds
evident in young H-rich (O-star type) CSPN, through which
$\sim$ 10$^{-8}$ to 10$^{-6}$ M$_\odot$ yr$^{-1}$ of mass can be driven
off the star, and with wind terminal velocities of up to a few 1000s
km s$^{-1}$ (e.g. Kudritzki, Urbaneja {\&} Puls, 2006, and references
therein). Accurate knowledge of the fast stellar winds is important for
understanding the evolution of CSPN and for decoding their spectral
diagnostics. Mass loss from CSPN is also a key process in the
interacting stellar wind (ISW) model that may explain some of the
PN morphologies observed (e.g. Balick {\&} Hajain 2004).
Stellar binarity could
also have a pivotal role in shaping non-spherical nebulae (see e.g.
De Marco 2009) and central star binarity
represents a useful potential
laboratory for examining asymmetric wind phenomenon.
In this respect PNs can contribute to our knowledge of analogous settings
such as Wolf-Rayet nebulae, luminous blue variables (LBVs) and
Galactic Centre phenomena.

Our focus in this paper is to explore similarities between the
fast wind properties of H-rich CSPN and the radiation pressure driven
winds of luminous OB stars, and in particular the evidence for
inhomogeneities and structure in the winds.
{\it International Ultraviolet Explorer (IUE)} spectroscopy of OB star
winds has revealed widespread variability seen primarily in the blueward
extended absorption troughs of resonance line doublets (e.g. Massa et al.
1995; Kaper et al. 1997; Prinja et al. 2002). Several case studies have
demonstrated that the UV line profile fluctuations in massive, luminous stars
are due to evolving (over days), organised structures in the wind, which
may have a physical origin rooted to the stellar surface
(e.g. Fullerton et al. 1997; de Jong et al. 2001).
The presence of structure and wind clumping affects mass-loss rates
derived from different spectral diagnostics (e.g.
Sundqvist, Puls {\&} Feldmeier 2010;
Prinja {\&} Massa 2010) and the resultant wind porosity can complicate
significantly the radiative
forcing used in theoretical mass loss rate determinations
(e.g. Muijres et al. 2011).
The fast winds of H-rich CSPN provide a laboratory for constraining
the origin of wind structure and comparisons to massive star outflows.
The CSPN offer unique tests for the role of radiative instabilities,
large scale structures, and stellar rotation over a wide range
of critical length scales. For example the rotation period
(for fixed projected rotation velocity) and wind flow time (for
fixed terminal velocity) both scale as stellar radius, and are therefore
reduced in CSPN relative to normal OB stars.
Our present study of structure in CSPN fast winds also has
potential relevance to studies of variable features
on the stellar surface that then propagate into the wind, such as
the sub-surface convection zones studied by Cantiello et al. (2009).

The identification and characterization of variable wind structure in hot stars
demands time-series spectroscopy of diagnostic spectral lines, over
an appropriately extended time-scale and with sufficiently intensive
monitoring. The {\it Far Ultraviolet Spectroscopic Explorer (FUSE)} time-series
of NGC~6543 studied by Prinja et al. (2007) is the most suitable dataset
currently available in the UV archives for the analysis of large-scale
wind structure in CSPN. Their analysis revealed clear evidence for
recurrent discrete absorption components (DACs) in NGC~6543 akin to
the features commonly seen in several OB stars. We aim in the present
paper to place the findings of Prinja et al. (2007) on a wider observational
footing and we present here a survey of {\it FUSE} time-series
datasets of CSPN.
We extract the time-variable FUV characteristics of
NGC~6826, IC~418, IC~4593 and IC~2149. An additional archival dataset
of NGC~6543 (to the fuller one presented by Prinja et al. 2007) is
also included here.
Other UV archives such {\it IUE} and {\it HST} do not currently contain
any additional high-resolution time-series datasets of H-rich CSPN
that are appropriate for studying variable wind structure.

In the following sections we introduce the {\it FUSE} time-series
samples and present an analysis of
temporal features in the Doppler-shifted absorption troughs.
Velocity measurements are supplemented by radial optical
depths derived from empirical SEI (Sobolev
with Exact Integration; Lamers et al. 1997; Massa et al. 2003)
line synthesis modelling of the P{\sc v} $\lambda\lambda$1118, 1128
resonance doublet.
The determination
of fundamental parameters is extended to include estimates of
the projected rotation velocity of the 5 target central stars using
the fast Fourier transform (FFT) technique.

\begin{figure}
 \includegraphics[scale=0.49]{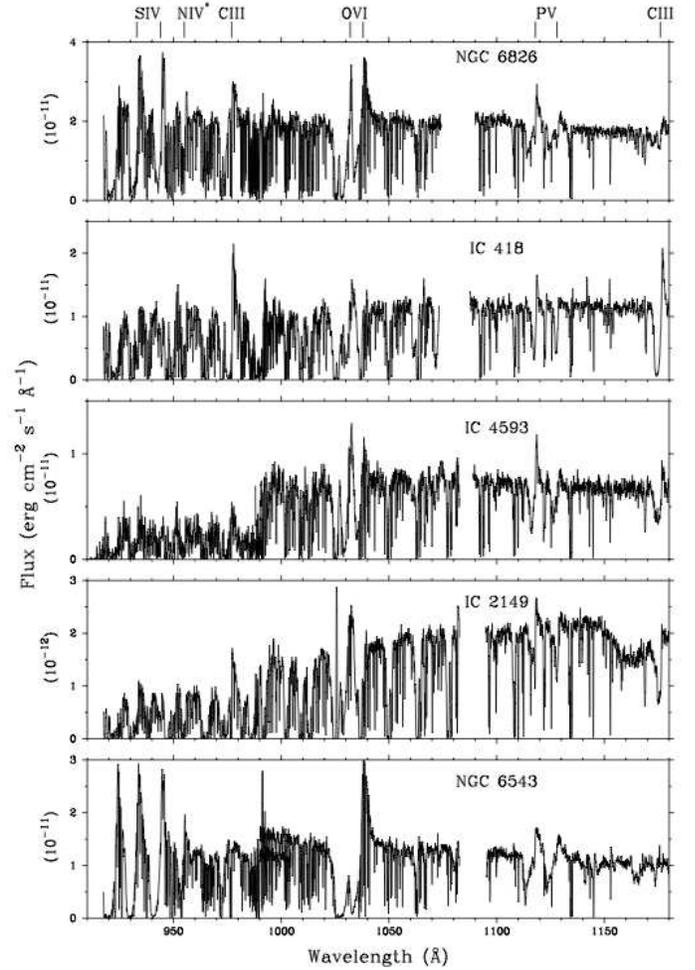}
 \caption{Mean {\it FUSE} spectra for the 5 CSPN stars examined
for wind variability; key wind-formed line of interest
are marked.}
\end{figure}

\begin{figure*}
 \includegraphics[scale=0.34]{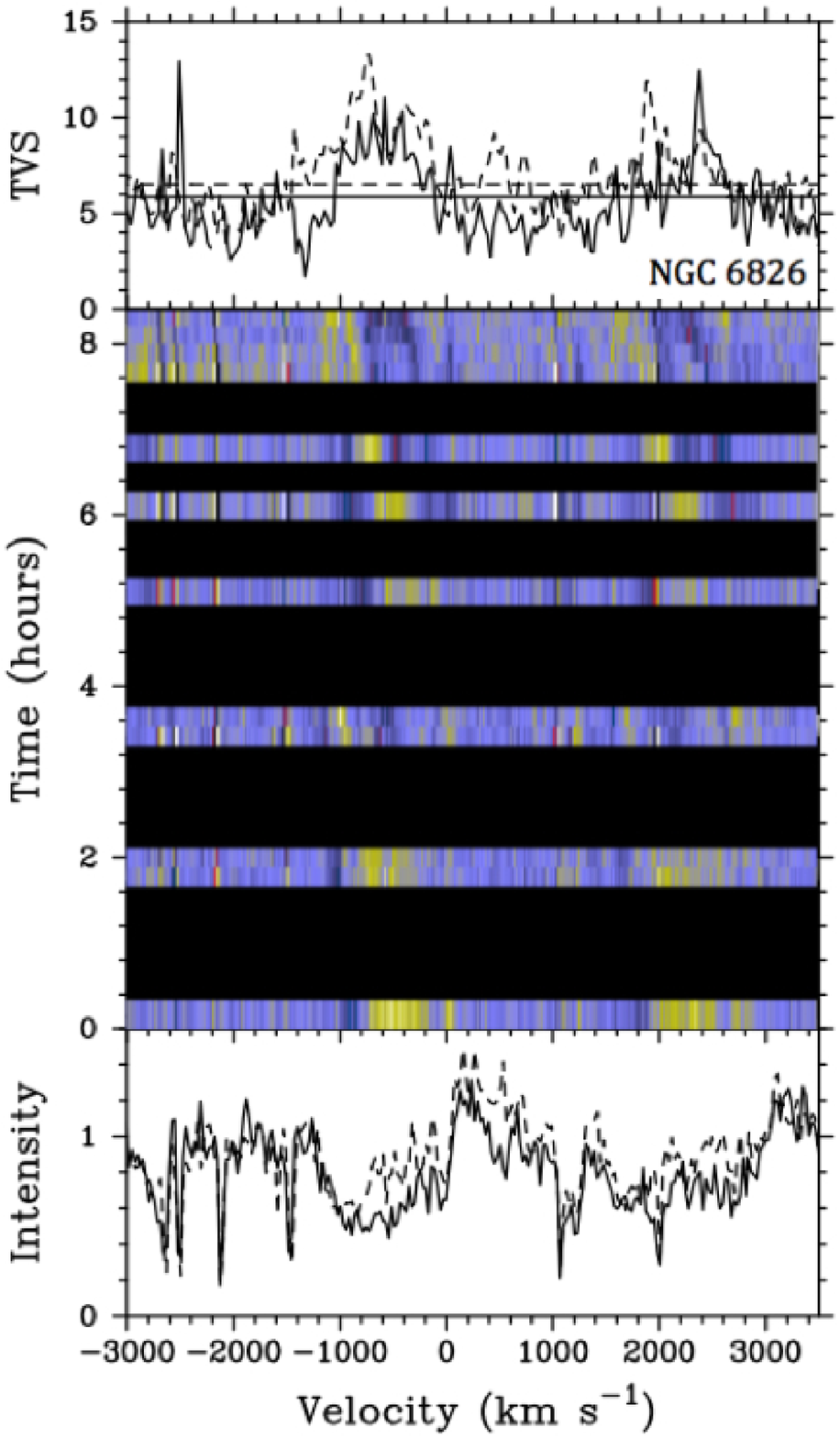}
 \includegraphics[scale=0.33]{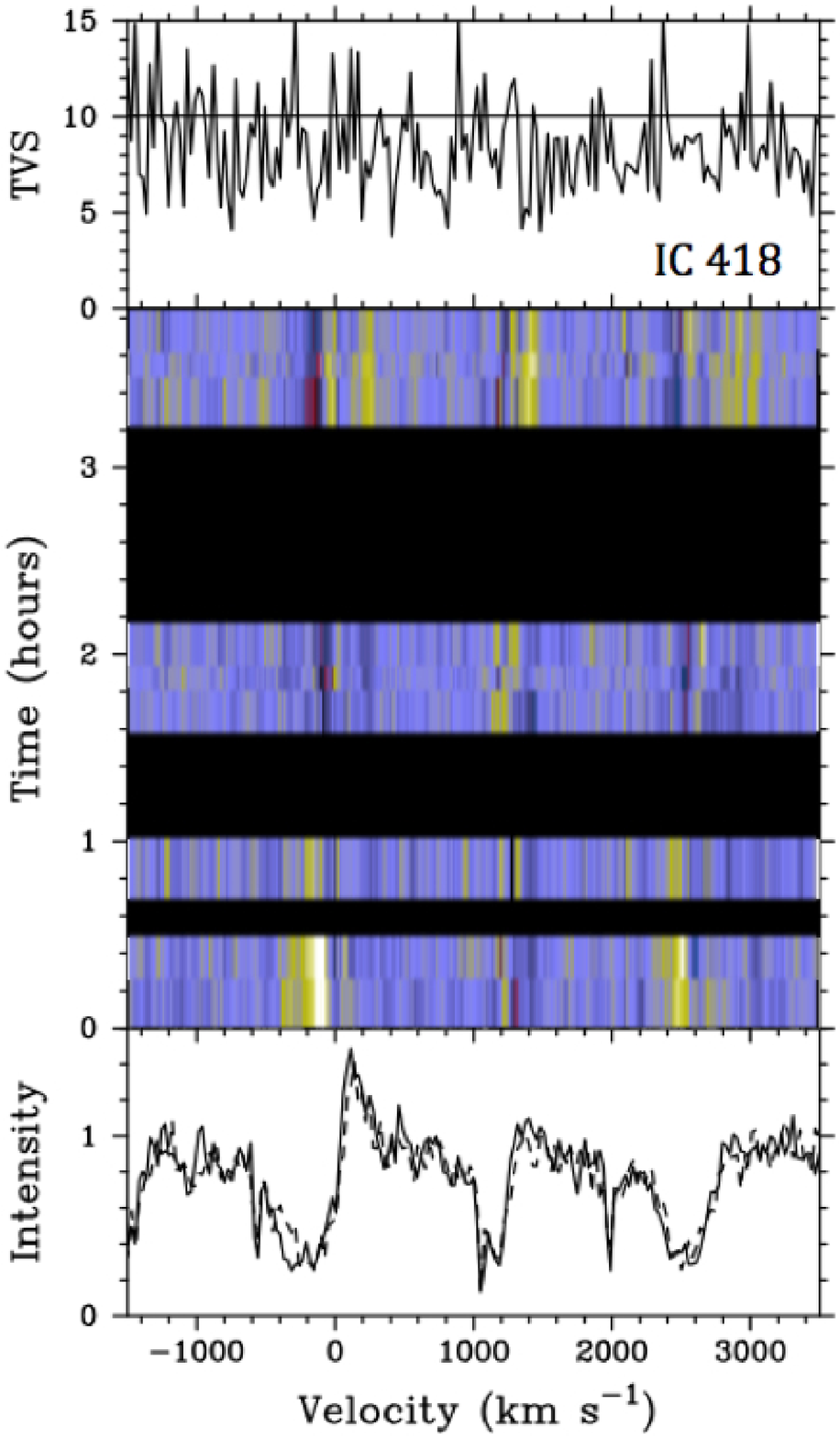}
 \includegraphics[scale=0.33]{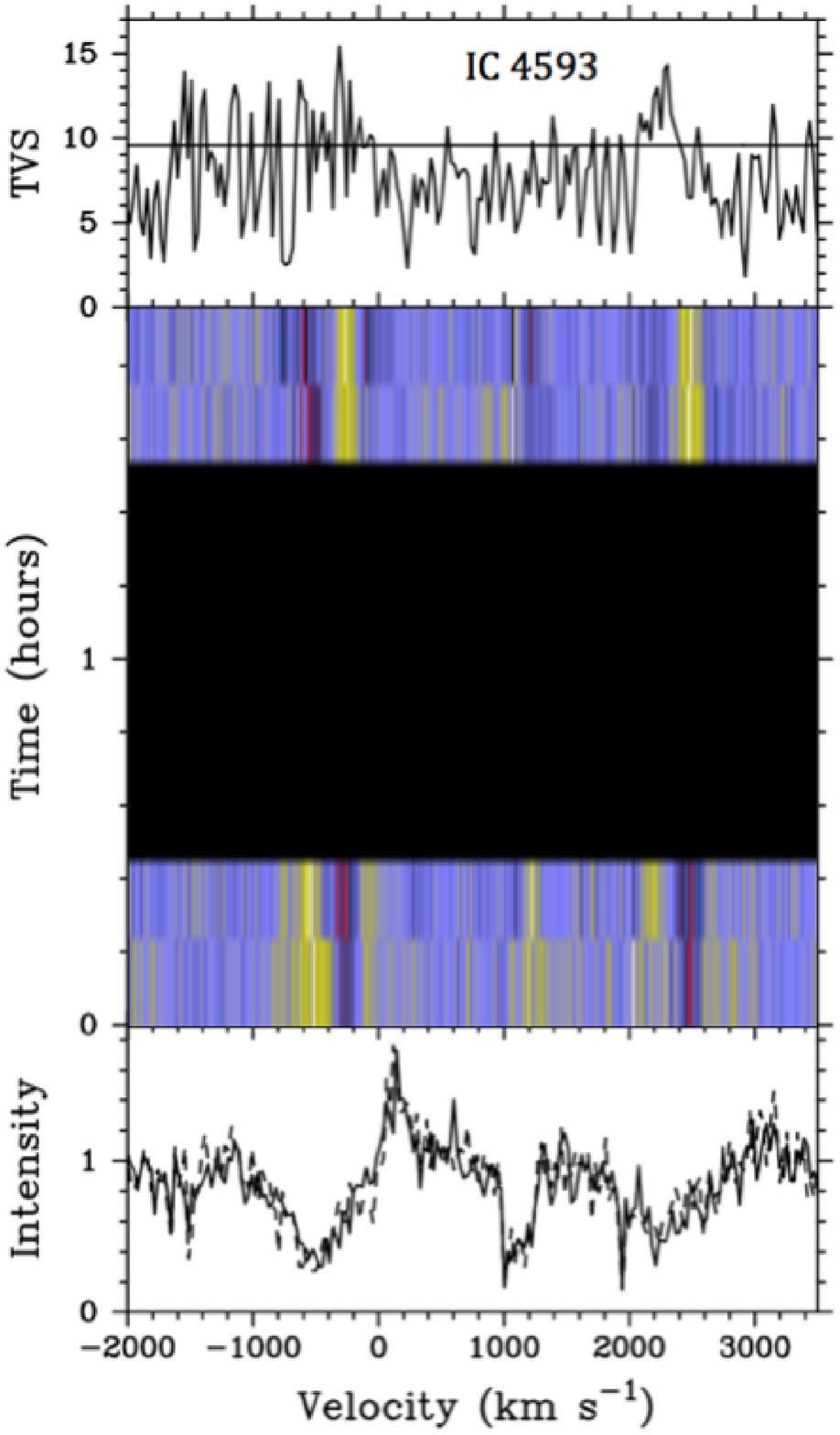}
 \includegraphics[scale=0.33]{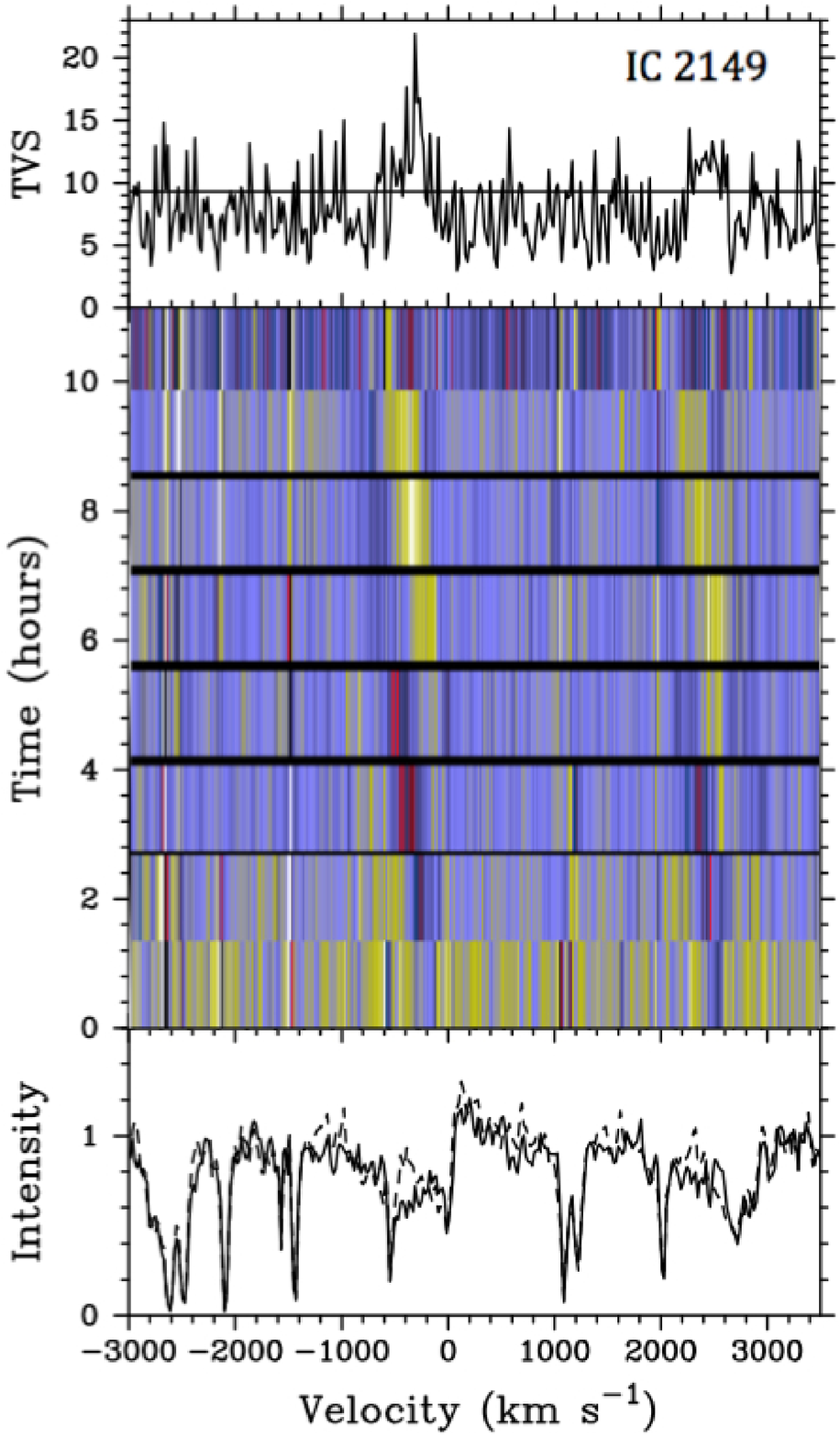}
 \includegraphics[scale=0.33]{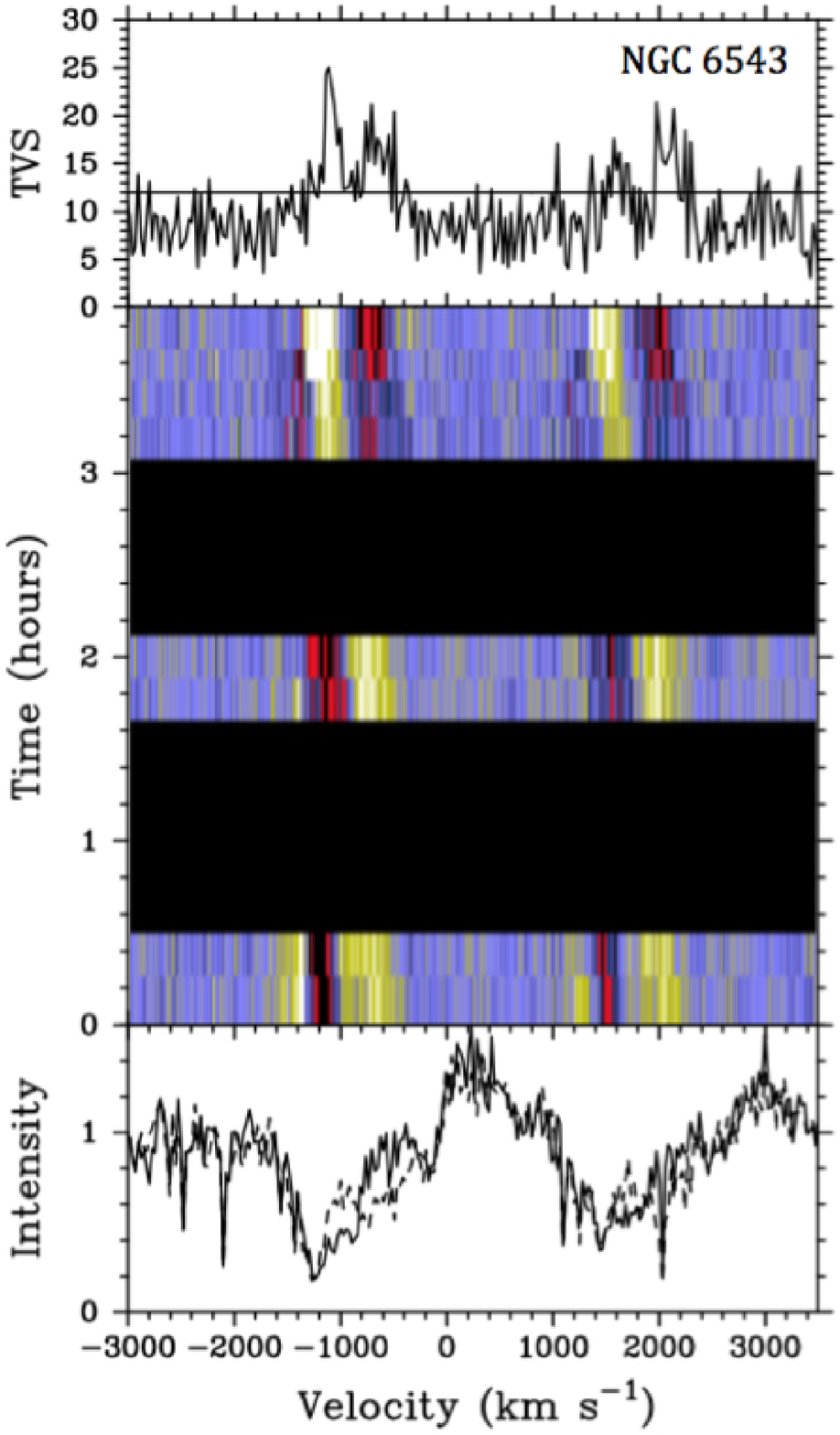}
 \includegraphics[scale=0.32]{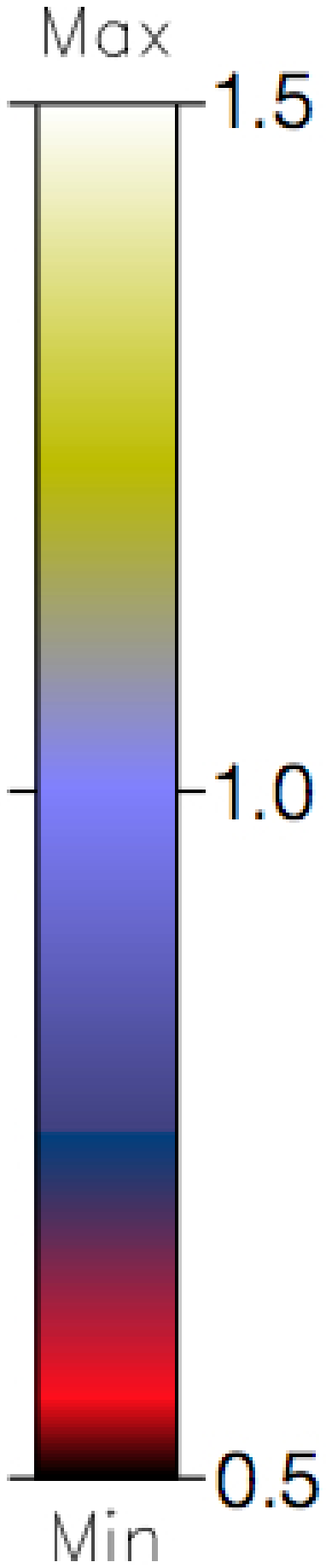}

 \caption{Dynamic spectra showing the evidence for
structure in P{\sc v} $\lambda\lambda$1117.98, 1128.01
in the form of 
discrete absorption components (DACS)
in the winds of PN central stars: Clockwise from the 
top-left, NGC~6826, IC~418,
IC~4593, NGC~6543 and  IC~2149.
For each case, the upper panel displays the
temporal variance spectrum (TVS) and 95{\%}
confidence limit (horizontal lines). The panels below
the dynamic spectra show pairs of the variable P{\sc v}
line profiles.
In the case of NGC~6826 the TVS for
an additional dataset from 2007 January is also shown
(dashed line).
}

\end{figure*}

\section{The {\it FUSE} time-series sample}

Our CSPN sample is derived from archival {\it FUSE} data available via the NASA
Multimission Archive at Space Telescope (MAST).
We initially searched for {\it FUSE} observations
in the object category 70 (`Planetary Nebula and central star').
The $\sim$ 200 spectra listed were then narrowed to H-rich (O-type) stars that
had at least one unsaturated, but well-developed P~Cygni line profile
in the {\it FUSE} range. From these data we only retained those objects that
had at least 4 sequential observations available within a span of
12 hours; this was deemed a minimum requirement to permit an inspection
for the incidence of organised wind structure features in the line profiles.
These criteria reduced the available sample to the 5 H-rich CSPN listed
in Table~1. (The epoch of the dataset of NGC~6543 included here predates
the time-series presented by Prinja et al. 2007.)
The datasets were secured using either the LWRS (30"$\times$30") or
MDRS (4"$\times$20") spectrograph apertures (Table 1).
In all cases, except IC~2149, the data were recorded in HIST mode,
with individual exposure times between $\sim$ 7 to 8 minutes.
The time-series of IC~2149 is based on temporal segments obtained
in time-tagged event lists (TTAG mode), with a total exposure of
21910 sec.
The individual spectra
were aligned using strong interstellar lines to account
for wavelength drifts in the {\it FUSE} non-guide channels.
The data studied here were gridded to a resolution of 0.1{\AA}
and the mean spectrum for the {\it FUSE} channels available are shown in
Fig. 1.
Adopted fundamental stellar parameters based on published
non-LTE model atmosphere analyses are collated in Table~1.

The typical continuum signal-to-noise of an individual {\it FUSE}
spectrum is only $\sim$ 15 and at this quality the most suitable spectral
line for a study of variable wind structure is the
P{\sc v} $\lambda\lambda$1118, 1128 resonance line doublet.
The P{\sc v} line is unsaturated and its doublet separation of $\sim$ 2690 km 
s$^{-1}$ is large
enough compared to the typical terminal velocity, $v_\infty$ of our targets
such that the each component is reasonably well resolved, thus helping
to remove ambiguities in the occurrence of optical depth and
velocity structure in limited time-series datasets.

All five targets listed in Table 1 have been previously documented by
Handler (2003) as exhibiting semi-regular photometric variability,
over time-scales ranging from $\sim$ 3.5 to 6.5 hours.
Handler (2003) proposes that these stars form part of a class of
variable objects labelled ZZ Leoparis stars, where the prototype is the
central star of IC~418 (HD~35914).
Previously, pairs of high-resolution
$IUE$ spectra of our target stars secured more than a year apart revealed
mostly tentative evidence for changes in the blue-ward wings of
saturated P~Cygni profiles (e.g. Patriarchi {\&} Perinotto, 1995, 1997).
We present here analyses of organised,
and temporal, stellar wind structure in this sample of H-rich CSPN.

\subsection{Optical spectra}

High resolution (R=68000) optical spectra of the stars in our sample were 
collected with the echelle spectrograph ESPaDOnS at the 3.6-m Canada-France-Hawai'i 
telescope
in the summit of Mauna Kea during the spring of 2008, under excellent weather
conditions. The collected data were processed with the data reduction package
Libre-ESpRIT (Donati et al. 1997). The extracted spectra cover
almost entirely the 3700--10500\AA~wavelength range, with a signal-to-noise ratio
in the continuum above 200, except for the very blue end of the observed range.
In the following, we will use the optical spectra to complement our FUV series in
order to estimate the rotational velocities of our stars.


\begin{table*}
 \centering
\caption{Fundamental central star parameters.}
  \begin{tabular}{lllllllll}
  \hline
CSPN & Sp.type & v$\sin i$ (km/s) & Distance (pc) & T$_{\rm eff}$ (kK) & Log $g$ (cgs) & L$_\star$/L$_\odot$ & 
He abun. & Ref. \\
\hline

NGC~6826 & Of (H-rich)  &  50  & 1590   & 46.0  & 3.80 & 4.11 & 0.10   & KUP2006  
\\ 
IC~418        & Of (H-rich)  &  56 & 1300  & 36.7  & 3.55 & 3.88 & 0.25   & MG2009    
\\  
IC~4593      & Of (H-rich)  &  55 & 3225  & 40.0  & 3.60 & 4.04 & 0.10   & KUP2006  
\\  
IC~2149      & Of (H-rich)  &  54 & 3258  & 38.0  & 3.45 & 4.45 & 0.10   & U2012    
\\   
NGC~6543 & Of (H-rich)  &  85 & 1623  & 67.0  & 5.30 & 3.76 & 0.10  &  G2008      
\\    
\hline
\end{tabular} \\
References: Kudritzki, Urbaneja \& Puls (2006; KUP2006); 
Morriset \& Georgiev (2009; MG2009); 
Georgiev et al. (2008; G2008); Urbaneja et al. 2012 in preparation (U2012).
 Distances taken from Stanghellini, Shaw \& Villaver (2008), except for
IC~418, taken from Guzman et al. (2009).
The He abundances are relative to H.
The projected rotation velocities (v$\sin i$) are derived
in Sect. 5.
\end{table*}



\begin{table*}
 \centering
\caption{FUSE fragmented time-series datasets.}
  \begin{tabular}{llllcc}
  \hline
CSPN & FUSE Prog. & Aperture & Date & No. of spectra & Total span (hrs) \\ 
\hline

NGC~6826 & P1930401 & MDRS & 7 Aug. 2000 & 12 & 8.5  \\
IC~418 & P1151111 & LWRS & 2 Dec. 2001 & 9 & 3.8  \\
IC~4593 & B0320102 & MDRS & 3 Aug. 2001 & 4 & 1.9  \\
IC~2149 & P1041401 & LWRS & 2 Dec. 1999 & 8 & 11.1  \\
NGC~6543 & Q1080202 & MDRS & 1 Oct. 2001 & 8 & 3.7 \\

\hline
\end{tabular} \\
\end{table*}


\begin{figure*}
 \includegraphics[scale=0.57]{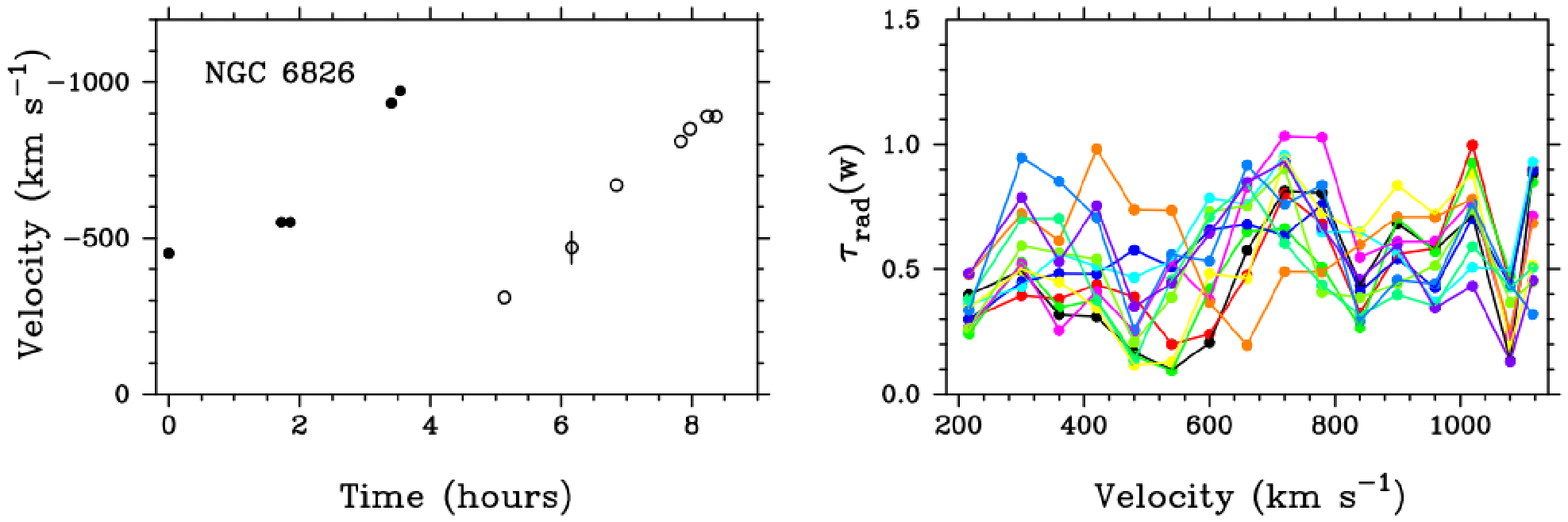}
 \includegraphics[scale=0.57]{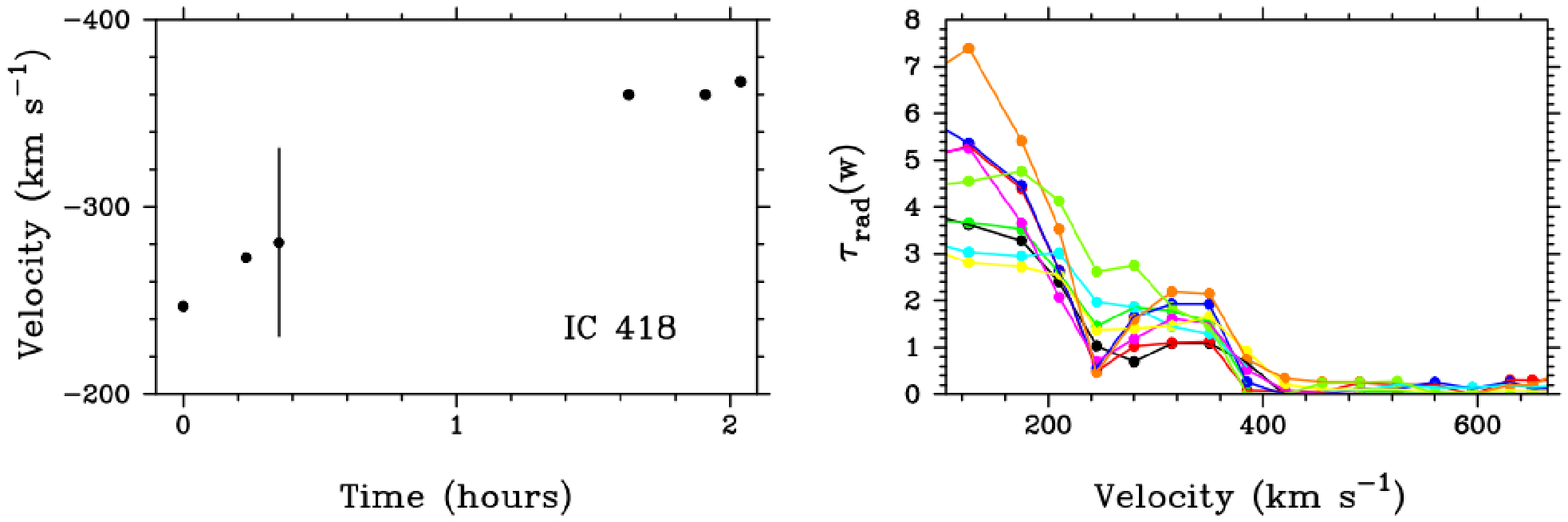}
 \includegraphics[scale=0.57]{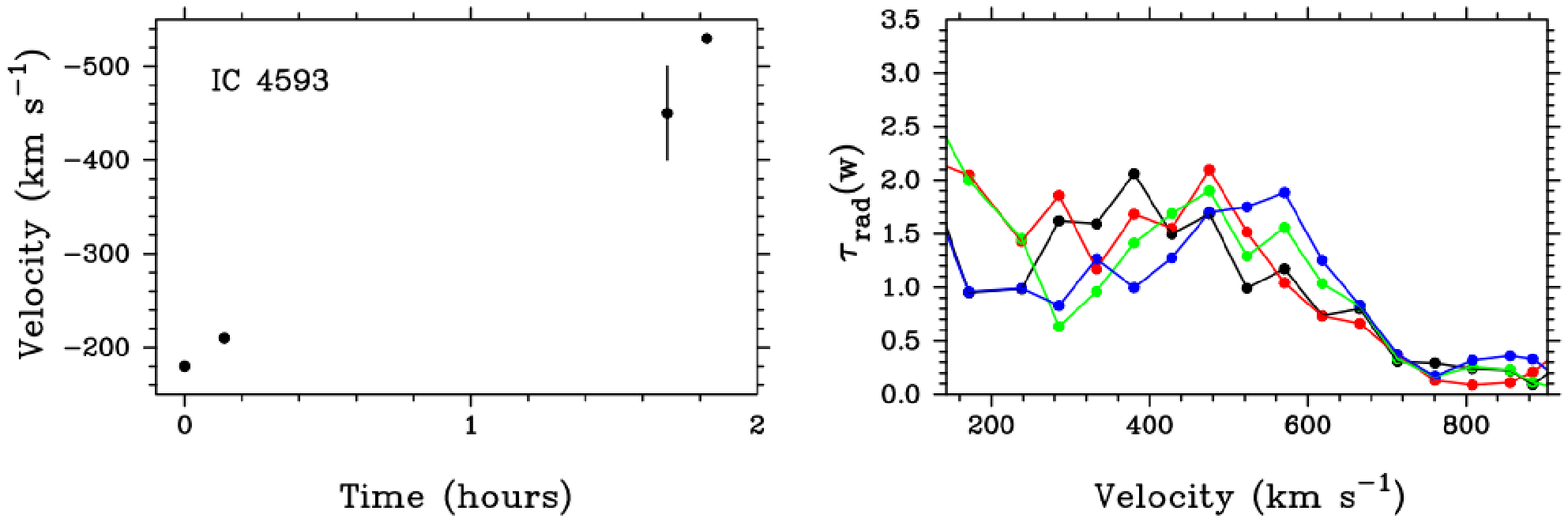}
 \includegraphics[scale=0.57]{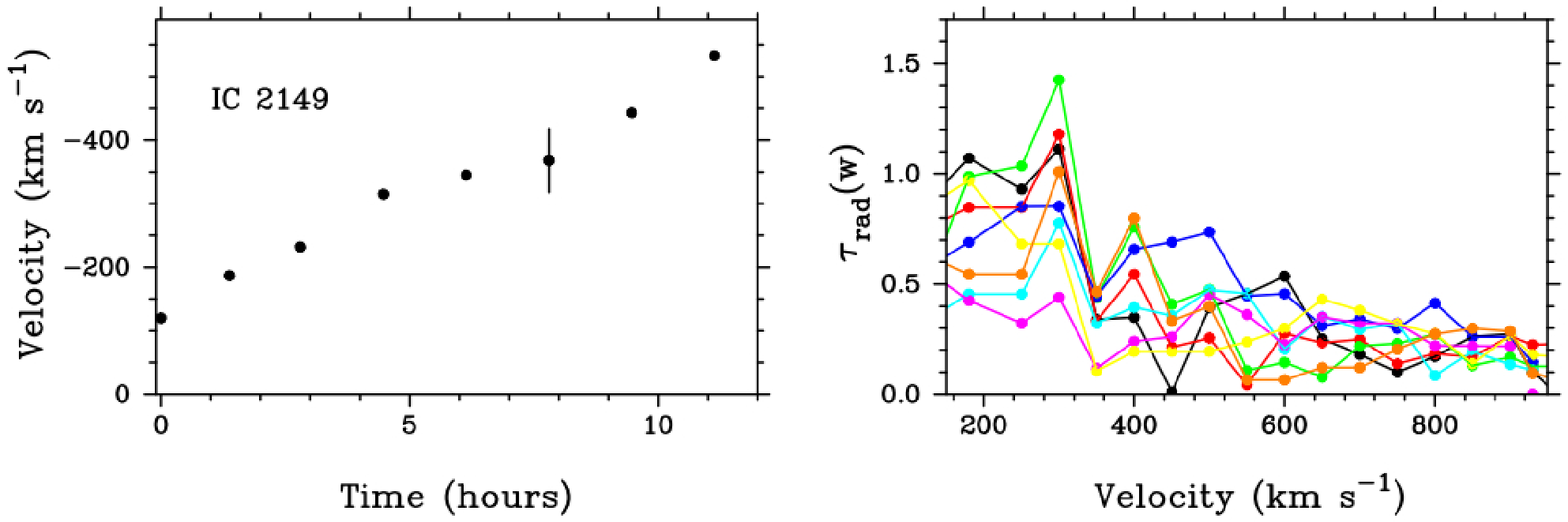}
 \includegraphics[scale=0.57]{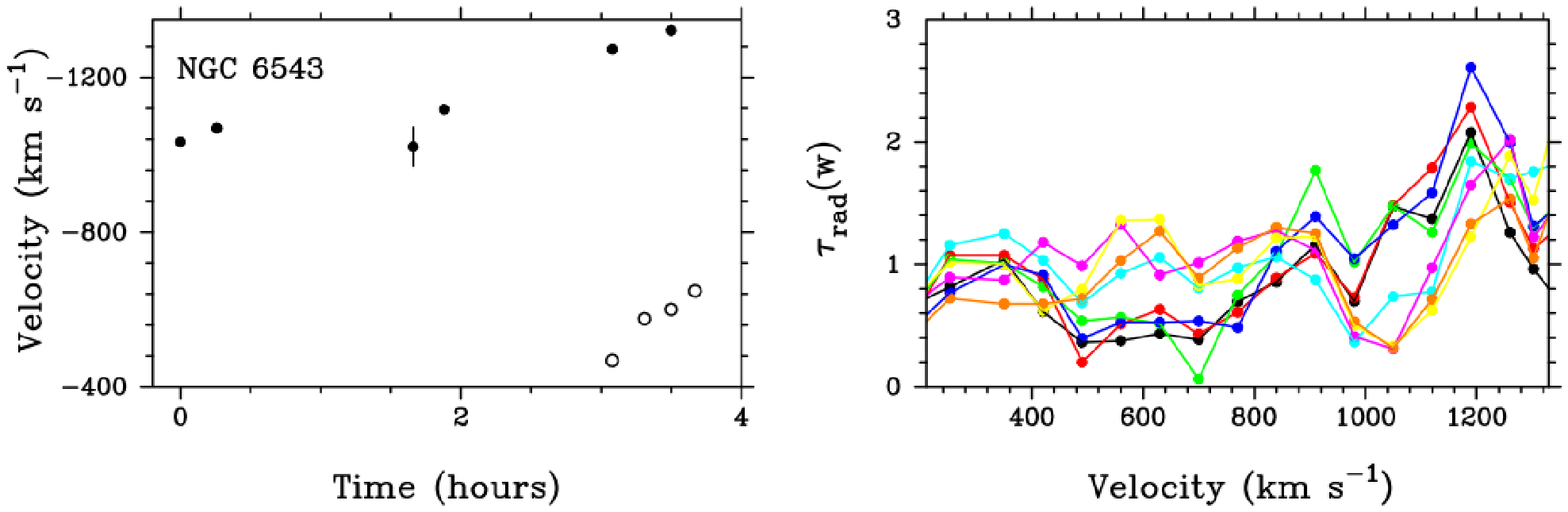}
 \caption{Left-hand panels: The central velocities of migrating discrete
absorption components (DACs) as a function of time
in the P{\sc v} doublet profiles of the target CSPN.
The corresponding changes in (SEI-derived) radial optical depth (Sect. 4) for
the time-series line profiles are shown in the (right-hand
panels).
We assign an error of $\pm$ 5{\%} in the $\tau_{\rm rad}(w)$
values adopted in each velocity bin.
}
\end{figure*}


\begin{table*}
 \centering
\caption{SEI line-synthesis results.}
  \begin{tabular}{lllllll}
  \hline
CSPN & a$_{\rm DAC}$ (km s$^{-2}$) & $v_\infty$ (km s$^{-1}$) & $\beta$ & $v_{\rm 
turb}$ &
range $\langle\mdot\,qP^{4+}\rangle$
& $\tau_{rad}$(max)/$\tau_{rad}$(min) \\
 & & & & & (10$^{-9}$ M$_\odot$ yr$^{-1}$) & [DAC]\\
\hline

NGC~6826 & 0.053 & 1200 & 1.2 & 0.04 & 6.7 $-$ 8.9 & 7.8 (at 0.45 $v_\infty$) \\
IC~418 & 0.016 & 700 & 3.0 & 0.13 & 3.5 $-$ 6.7  & 5.7  (at 0.35 $v_\infty$) \\
IC~4593 & 0.042 & 950 & 3.0 & 0.07 & 14.3 $-$ 16.2 & 2.9 (at 0.30 $v_\infty$) \\
IC~2149 & 0.009 & 1000 & 3.0 & 0.09 & 8.5 $-$ 12.3  & 10.5 (at 0.55 $v_\infty$) \\
NGC~6543 & 0.022 & 1400 & 1.0 & 0.05 & 4.6 $-$ 5.9 & 4.8 (at 0.35 $v_\infty$) \\

\hline
\end{tabular} \\
\end{table*}


\section{Discrete absorption components}
Inspection of the line profiles in the {\it FUSE} fragmented time-series
datasets listed in Table~1 suggests that all five target CSPN have outflows
that vary on hourly timescales. The fluctuations are mostly
at the $\sim$ 10 to 15{\%} of the continuum level and over a substantial
portion of the absorption trough in P{\sc v}, C{\sc iii} and
O{\sc vi}. Corresponding changes are seen in different line species,
which is more consistent with a notion of line-of-sight density
changes rather than extreme ionization shifts.

Following a well established approach, dynamic spectra (colour-scale
images) of the time-series data of each CSPN are presented in Fig. 2.
Ratios of individual P{\sc v} $\lambda\lambda$1118, 1128 line
profiles divided by the mean for the time-series are used to
highlight the variability present. For each dynamic spectrum in
Fig. 2, the uppermost-panel shows the temporal variance spectrum,
TVS (see e.g. Fullerton et al. 1996), and the corresponding bottom
panel exhibits pairs of P{\sc v} line profiles to illustrate the
typical intensity changes.
The TVS estimates the significance of the line profile
variability as a function of velocity.
Not only do the central stars reveal $\sim$ hourly variability in the
fast wind, the evidence in Fig. 2 is that the changes are primarily
in the form of blue-ward migrating features, closely matching in
empirical appearance the discrete absorption components (DACs)
that are commonly seen in the UV spectra of OB stars (see Sect. 1).
We note also the ubiquity of the variability in CSPN such that
whenever the spectral diagnostics and data are available,
variability is detected.
The results in Fig. 2 indicate wind changes occurring at
very low velocities (less than 100 km s$^{-1}$) in some
of the time-series, which (as in OB stars) suggests a
causal connection to deep-seated structure at or close to
the stellar surface.
Brief summaries of the DACs in each CSPN follow below:

{\it NGC~6826} $-$ We identify the occurrence of two sequential
DACs over $\sim$ 8.5 hours. The first feature is seen already
formed at the start of the time-series at $\sim$ $-$450 km s$^{-1}$
and migrates to $\sim$ $-$1000 km s$^{-1}$ over about 4 hours.
The second DAC is evident immediately after the data gap
between $\sim$ 4 to 5 hours in Fig. 2. This second feature
drifts from $\sim$ $-$250 km s$^{-1}$ to
$\sim$ $-$900 km s$^{-1}$ over $\sim$ 3.5 hours.

{\it IC~418} $-$
Despite the sparse dataset, and with the caveat that the variability is
weak,
we identify one DAC-like feature in IC~418.
It progresses from $\sim$ $-$250 km s$^{-1}$
at the start of the observations and has shifted to
$\sim$ $-$350 km s$^{-1}$ over the subsequent 2 hours.

{\it IC~4593} $-$ This is the most limited {\it FUSE} dataset for
our sample of stars, with only 4 spectra spectra spanning about
2 hours. Nevertheless there is evidence from the well separated
P{\sc v} doublets of temporal structure in the absorption trough.
We identify in Fig. 2 a feature that travels from
$\sim$ $-$200 km s$^{-1}$ to $\sim$ $-$500 km s$^{-1}$.
In this case, and in IC~418 above, confidence in this
detection comes from the fact that corresponding
structures are seen in both components of the P{\sc v} doublet,
and that similar behaviour is evident in for example
C{\sc iii} $\lambda$1176.

{\it IC~2149} $-$ One well monitored DAC is present in P{\sc v} profiles
of this central star. Using eight exposures, we trace its progression
from $\sim$ $-$100 km s$^{-1}$ to almost $\sim$ $-$500 km s$^{-1}$
over $\sim$ 11 hours.

{\it NGC~6543} $-$ Prinja et al. (2007) have already demonstrated
that NGC~6543 exhibits recurrent DACs in its wind-formed FUV lines,
using {\it FUSE} data secured in 2007 January. We show here (Fig. 2)
that two successive features are also present in data taken more than
5 years earlier. The phenomenon is therefore clearly long lived in
CSPN, as in the case for DACs in individual luminous OB stars.

\begin{figure*}
 \includegraphics[scale=0.39]{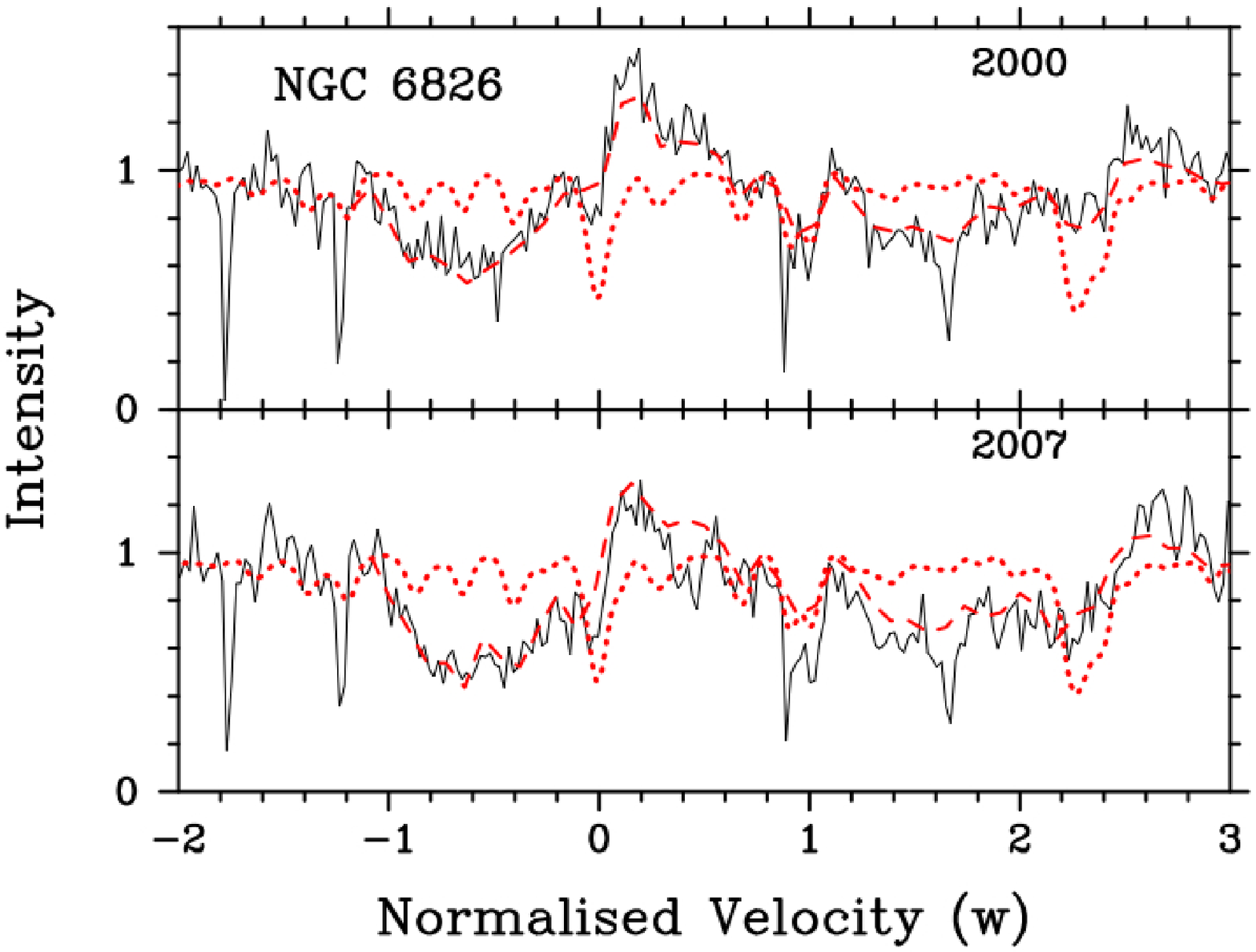}
 \includegraphics[scale=0.39]{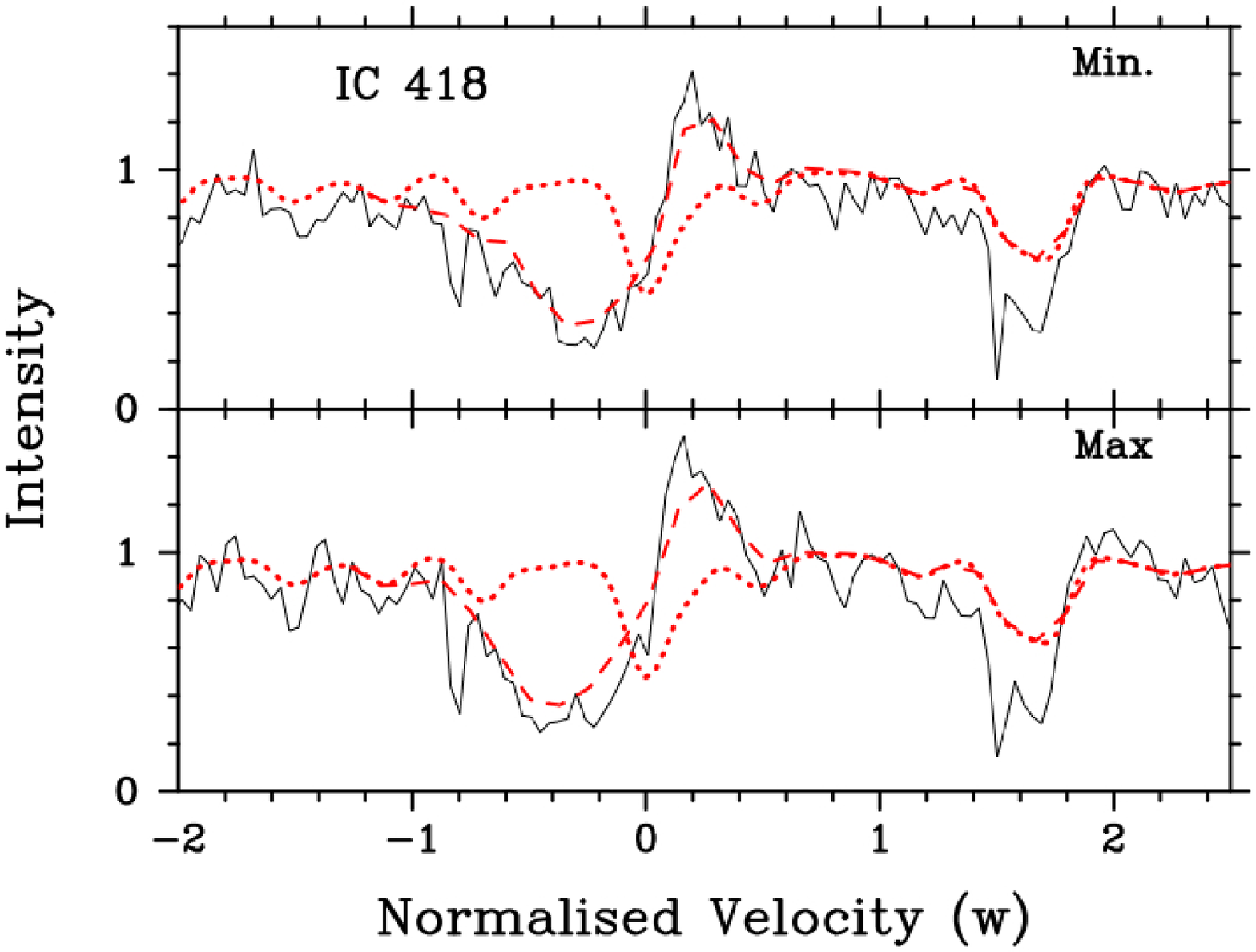}
 \includegraphics[scale=0.39]{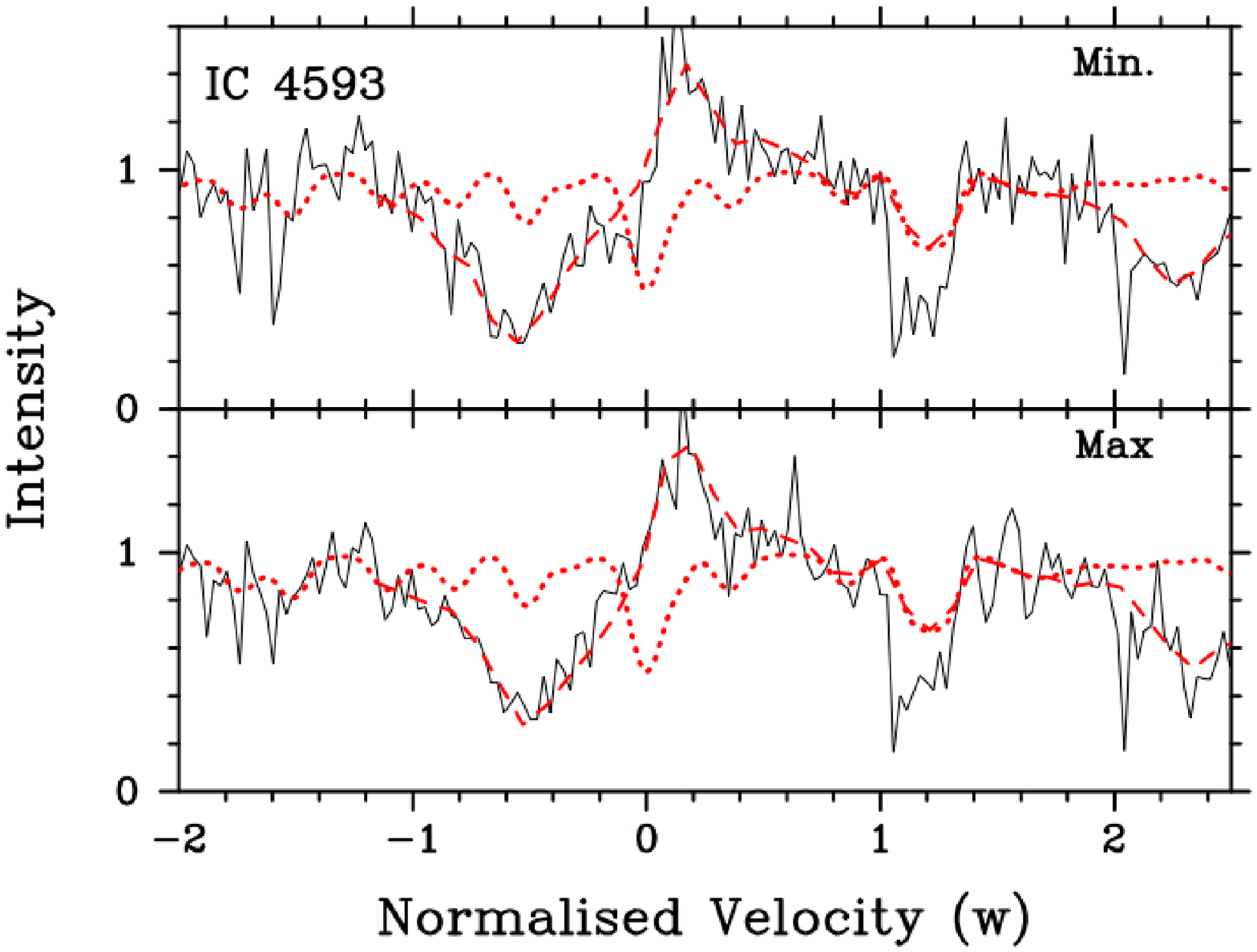}
 \includegraphics[scale=0.39]{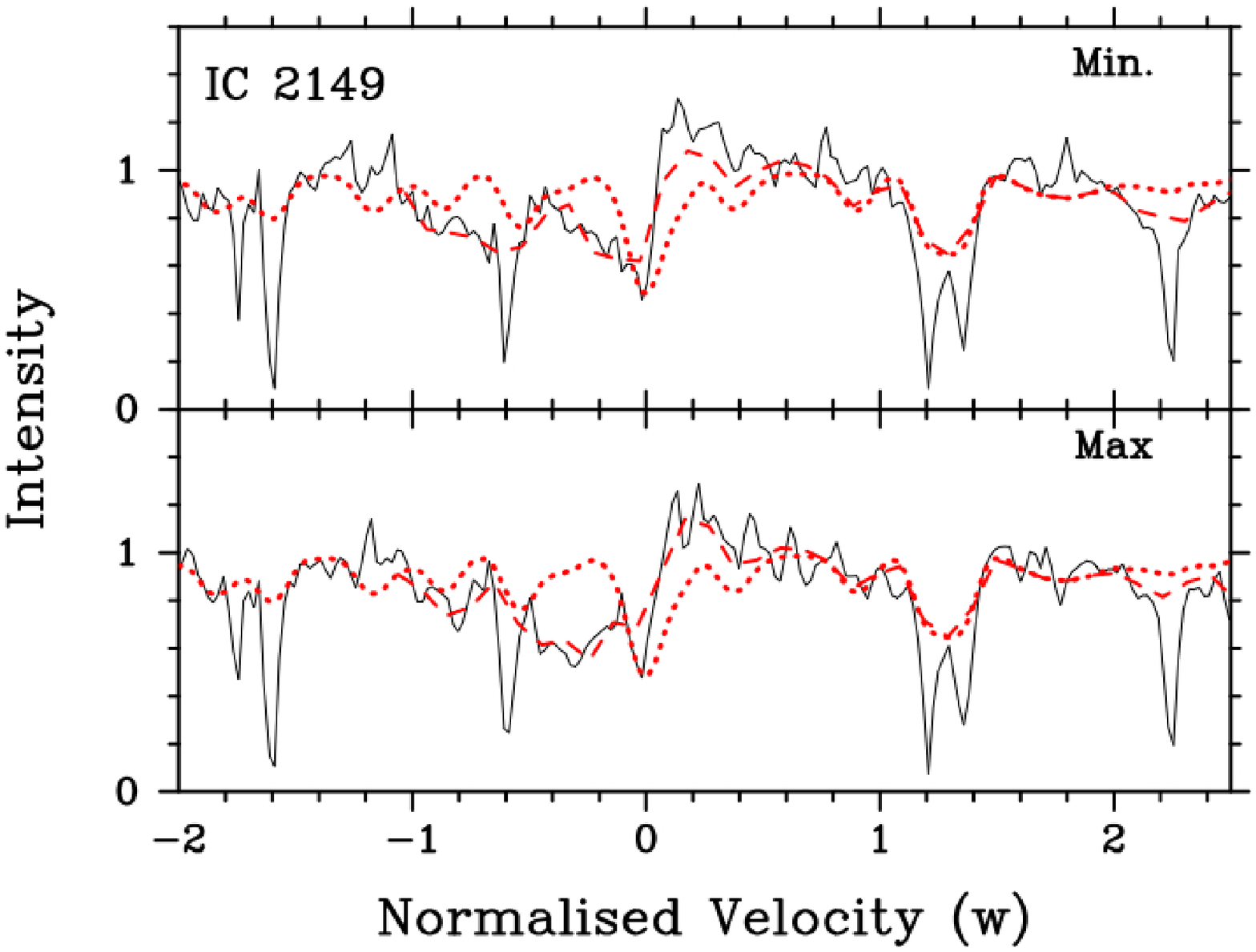}
 \includegraphics[scale=0.39]{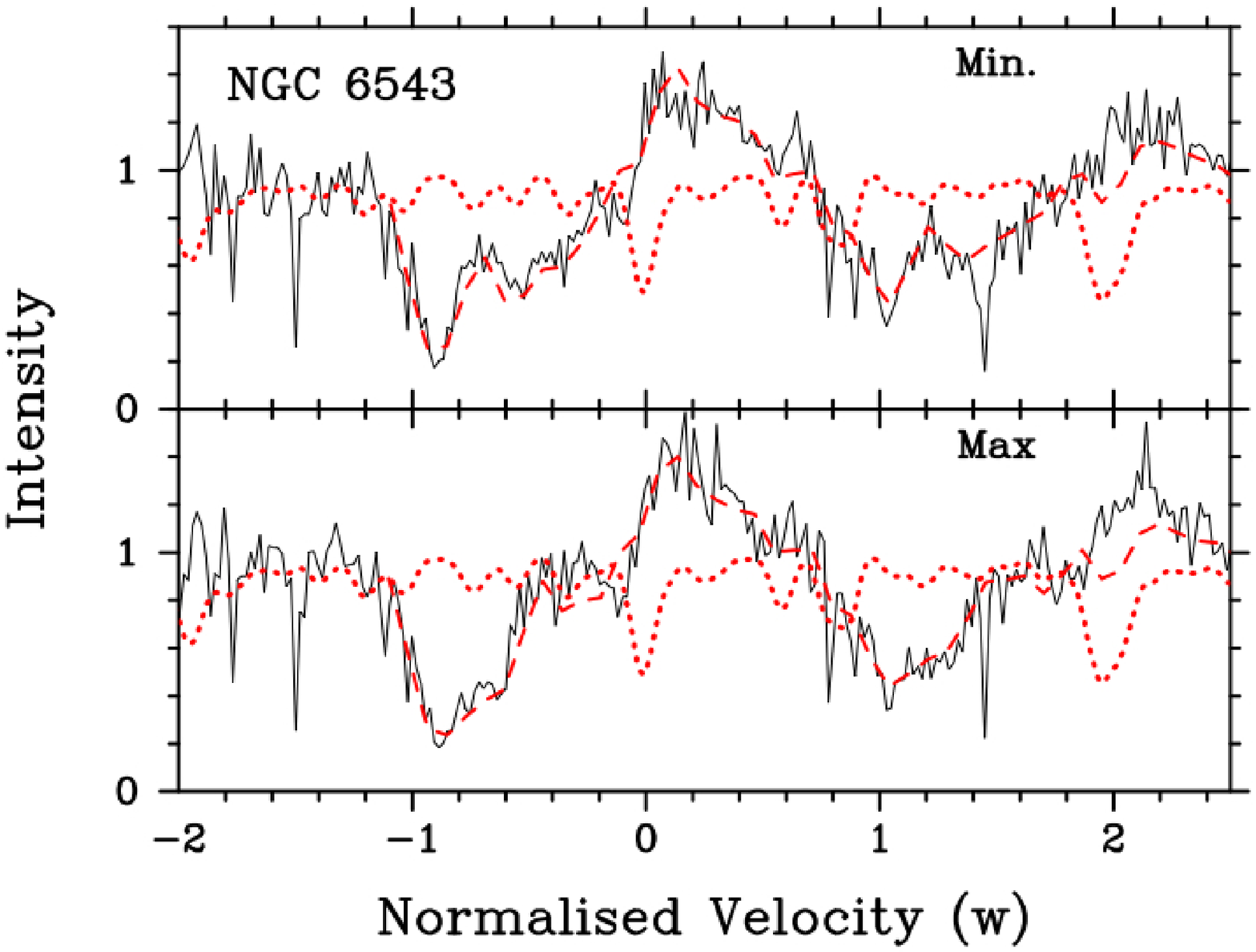}
 \caption{Examples of SEI-line synthesis model profile
fits (dashed line) to extreme variability cases in
P{\sc v} $\lambda\lambda$1117.98, 1128.01 for each
of the target stars. The adopted TLUSTY photospheric
spectrum is also shown in each PN central star case (dotted line)
For NGC~6826 an additional extreme case from 2007 January is
also matched (e.g. Fig. 2) (Normalised velocity = $v/v_\infty$).}
\end{figure*}

Guided by the dynamic spectra in Fig. 2, the central velocities
of discrete absorption components were estimated using by-eye
cursor values for each individual spectrum; the poor signal-to-noise
of the spectra precludes the use of any more sophisticated approach
(such as Gaussian fitting). We estimate that the DAC central velocities
quoted have errors of $\pm$ 50 km s$^{-1}$.
The velocities for individual DAC episodes in our sample are
plotted in Fig. 3 (left-hand panel) as a function of time.
Linear fits to these data yield accelerations which range between
$\sim$ 1.0 $\times$ 10$^{-2}$ km s$^{-2}$
to 5.3 $\times$ 10$^{-2}$ km s$^{-2}$.
The accelerations (a$_{\rm DAC}$) are listed in Table~3.

In the well monitored cases of NGC~6826, IC~2149 and NGC~6543
 a key characteristic of the DACs is that both their
propagation and recurrence time-scales (of $\sim$ few hours)
are longer than the characteristic wind flow time,
$R_\star/v_\infty$ ($\sim$ less than 1 hr).
The same relative comparison of timescales is commonly noted for
DACs evolving in the winds of OB stars and has led to the conclusion
in massive star studies that the large-scale structures mostly arise
due to perturbations, such as velocity plateaus, in the stellar wind
(see e.g. the hydrodynamical simulations of Lobel {\&} Blomme, 2008). 
Further comparisons of CSPN and O star DAC accelerations is
postponed to Sect. 6.

\section{Measurements from empirical line-synthesis fitting}
We seek in this section to derive additional empirical measures
that reflect the variability of the CSPN fast winds, with the
underlying premise of Sect. 2 that the line profile changes are
principally due to the actions of DAC-like features.
Our approach is described in Prinja et al. (2007) and relies on
matching the P{\sc v} doublet lines with model profiles generated
using a modified version the SEI methods of
Lamers, Cerruti-Sola {\&} Perinotto (1987).
The primary difference in the SEI code used here has been documented by
Massa et al. (2003) and the modification is motivated
by the need to reproduce UV line
profiles where the absorption troughs are not smooth, but
affected by localised optical depth structure.
Aside from all the standard parameterisations in the SEI code
(e.g. velocity law and macro-turbulence), the radial optical
depth, $\tau_{\rm rad}(w)$, is obtained from the fits by varying
21 independent velocity bins, each $\sim$ 0.05 $v_\infty$ wide.
This approach affords greater flexibility in matching the absorption
profile morphology, and in the current application offers a measure
of the change in optical depth due to the presence of DACs.

TLUSTY plane-parallel model atmosphere spectra (e.g. Hubeny {\&}
Lanz, 1995) were adopted as good approximations for the
photospheric contributions underlying the wind-formed
P{\sc v} line profiles in our sample stars.
The input photospheric absorption was included as
a 'lower boundary' in the SEI models, and we adopted TLUSTY
synthesised spectra corresponding closely to the T$_{\rm eff}$
and log $g$ values listed in Table~1 (and Sect. 5). Individual
P{\sc v} profiles were fitted by varying the 21
$\tau_{\rm rad}(w)$ bins and an overall measure of the
strength of the wind profile is provided by the
product of mass-loss rate and P$^{4+}$ ionization fraction
($\mdot\,q(P^{4+})(w))$.
Mean $\langle\mdot\,q(P^{4+})\rangle$ values over
0.2 $\le$ $v/v_\infty$ $\le$ 0.9 were determined, assuming solar abundances

We show in Fig. 4 the SEI line synthesis matches obtained for
cases of maximum and minimum values of $\langle\mdot\,q(P^{4+})\rangle$,
derived from fitting all the P{\sc v} profiles in the fragmented
time-series of each CSPN. The full range of
$\langle\mdot\,q(P^{4+})\rangle$ values for each central star
are listed in Table~3, together with the key SEI velocity law
parameters.
Even within a short dataset such as IC~418, the
$\langle\mdot\,q(P^{4+})\rangle$ values can differ by a factor
of 2 in less than 4 hours. The rest of the stars exhibit
$\sim$ 25{\%} changes in this measure over several hours.

The radial optical depth, $\tau_{\rm rad}(w)$, determined for
each P{\sc v} spectrum in our sample of CSPN is plotted as function
of normalised velocity on the right-hand panels in Fig. 2.
The changes in the optical depth arise from variability that is
principally due to the occurrence and temporal behaviour 
of the DACS (e.g. Fig. 2). The optical depth changes are
more substantial at low to intermediate velocities
($\sim$ 0.2 to 0.5 $v_\infty$), even in cases where the
P{\sc v} absorption troughs remain reasonably shallow and are
sensitive to line profile changes beyond 0.5 $v_\infty$.
The ratio of maximum to minimum radial optical depth measured
for each star is listed in Table 3.
The $\tau_{\rm rad}(max)$/$\tau_{\rm rad}(min)$
values correspond to specific velocities where a DAC is present.
This ratio rises to between $\sim$ 5 to 10 in our sample.
Assuming the line flux changes depend on the optical depth of
optically thick
large scale wind structures, and forward-scattering is small,
the results in Table 3 and Fig. 4 suggest that the DAC features
can occult between 10{\%} to 85{\%} of the stellar surface.
 
\section{CSPN rotational velocities}

There is a general dearth of published rotation velocity
measurements for CSPN. As is typical for hot, high gravity stars,
the H and He photospheric lines in CSPN are intrinsically very broad and
not ideally suited for determining the projected
rotation velocity, v$\sin i$.
From the perspective of understanding plausible physical mechanisms for 
generating large-scale structure in line-driven winds, the
parameter space offered by CSPN can be potentially valuable.
Stellar evolution calculations predict that single star
progenitors of CSPN should be slowly rotating
($V_{\rm eq}$ $<$ 1 km s$^{-1}$; see e.g.
Langer et al. 1999; Suijs et al. 2008).
Rotation velocity measurements of CSPN can therefore also
test the angular momentum transport mechanisms implemented
in the evolutionary calculations or alternatively highlight
implications for spin up due to binary interactions.

In order to estimate the projected rotational velocities of the stars, we 
consider a sample of FUV lines that are formed in or near the photosphere 
of these objects and that our models predict do not show a 
significant blend with other species, namely O~{\sc iii} 1151 and
O~{\sc iii} 5592. The results derived from these two lines can be validated with 
several other metal lines, but the specific lines depend on each star, 
given the range of physical parameters cover by our sample. Therefore, we 
primarily rely on the two O~{\sc iii} lines.

To derive a first estimation, we apply the FFT technique (Gray 1992). When
comparing the observed profiles with synthetic ones convolved with the derived
rotational rates, it is apparent that rotation alone does not account for all the
observed width
of the lines. By analogy with recent work on massive stars (see
Sim{\'o}n-D{\'i}az \& Herrero 2007 and references therein), we
consider that this extra broadening is caused by the presence of a macroturbulent
velocity field. While its interpretation is far from clear (although see Aerts et
al. 2009 and Sim{\'o}n-D{\'i}az et al. 2010 for
interpretations in term of stellar pulsations), is seems that the consideration of
macroturbulence (accounted for in the form of radial-tangential macroturbulence, 
Gray 1992) improves the profile fits.  Hence we apply a method to estimate this
extra broadening based on the construction of Gaussian profiles with
very narrow full width half maximum and with the same equivalent width
as the observed lines. These synthetic lines are first
degraded to the
spectral resolution of our observed data. 
The lines are then convolved with
the rotational velocities obtained with the FFT method. The best estimate of the
extra broadening is then found through a minimization of the residuals between the
observed and synthetic profiles convolved with a set of macroturbulent velocities.
In general, a marginally improved fit  is achieved when a
macroturbulence-to-vsini ratio of ~0.9-1.0 is considered.
A representative example of the FFT fit obtained for IC~2149
based on UV and optical spectral lines is shown in 
Fig. 5

\begin{figure*}
 \includegraphics[scale=0.79]{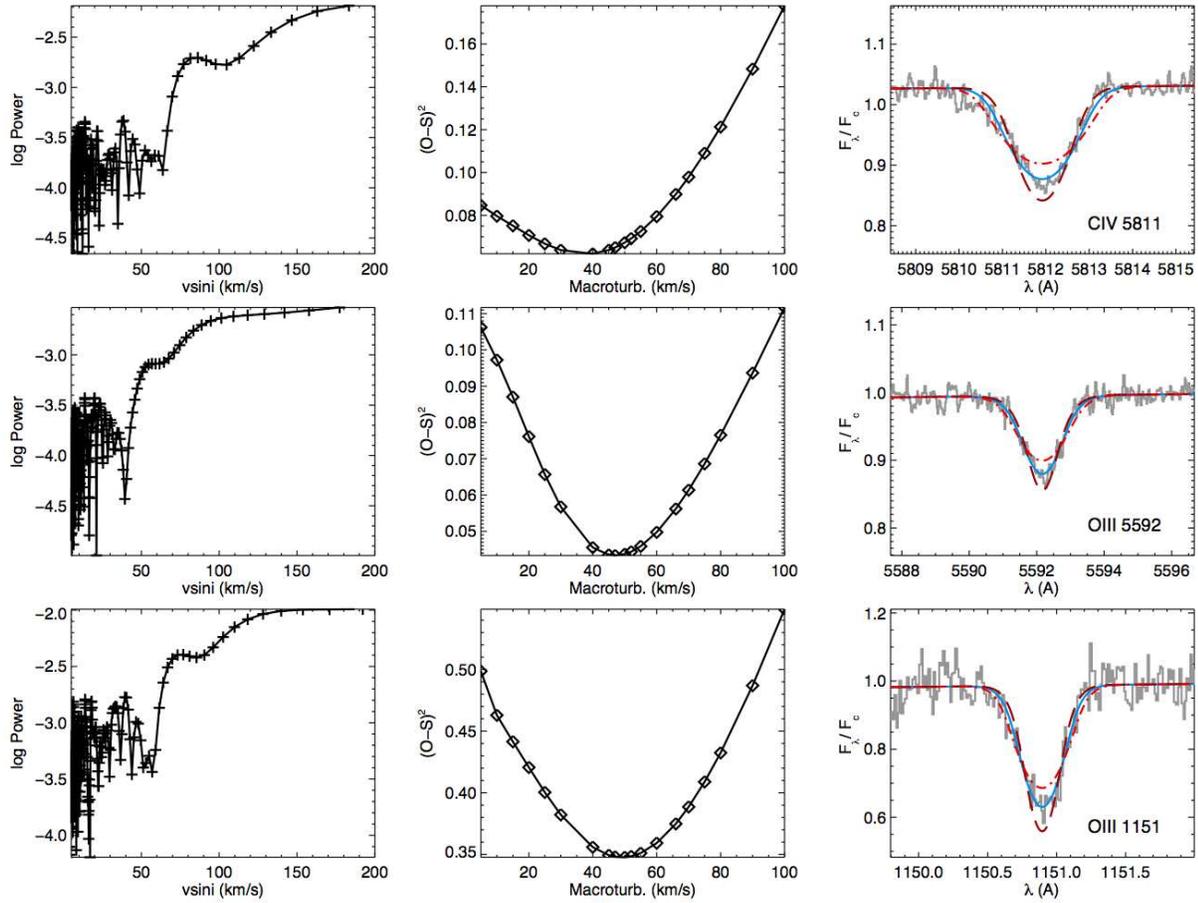}
 \caption{
Determination of the projected rotational velocity for the representative      
case of IC2149 based on UV and optical spectra lines.
For each line, the first panel shows the      
FFT determination of v$\sin i$; the second panel displays the 
determination of
the macroturbulence velocity, whilst the third panel shows a comparison of      
the observed (grey) and theoretical profile (solid)
convolved with a rotational profile characterized by the derived v$\sin i$      
and with a radial-tangential macroturbulence                                    
profile defined by the macroturbulence velocity. Two other synthetic            
profiles, convolved with rotational profiles for v$\sin                         
i\,\pm\,$15 km/s, are also shown (dashed, dot-dashed).
}
\end{figure*}

Our finally adopted v$\sin i$ values
obtained by means of the FFT technique, are quoted in Table 1.
We estimate conservative uncertainties of 10-15 km s$^{-1}$ in
these measurements.
The results and parameters listed in Table 1 indicate that the
maximum rotation periods for the central stars in our
sample range between $\sim$ 0.4 days (NGC~6543) and $\sim$
2.0 days (IC~418, IC~4593). 
Note that a substantial evolution in the
velocity of the DACs in witnessed during a fraction of the
maximum rotation periods, even in the case of NGC~6543.

\section{Discussion}
In the winds of luminous OB stars, which like CSPN are also radiation 
pressure-driven, the principal tracer of large-scale wind structure is the
incidence of recurrent discrete absorption components (DACs), which
migrate blueward across the absorption troughs of extended P~Cygni
line profiles. The DACs in OB stars have been extensively studied
by utilising high-quality UV time-series datasets obtained with the
$IUE$ satellite
(see e.g. Kaper et al. 1997; Massa et al. 1995; Prinja et al. 2002).
A key constraint on the origin of DACs comes from evidence that the
characteristic timescale for variability in OB star winds is
related to the rotation period of the star.
The measured acceleration and recurrence timescales of the DACs
reveal trends as a function of projected rotation velocity,
v$\sin i$, such that faster developing, more frequently
recurring DACs are
generally apparent in stars with higher v$\sin i$
(e.g. Prinja 1988; Kaper et al. 1997; Prinja et al. 2002).
Additionally studies of recombination formed Balmer lines suggest
that the cyclical behaviour revealed by the DACs extends down to
the inner wind, and possibly to the surface of the star
(e.g. de Jong et al. 2001).
Subsequent hydrodynamic models showed that coherent wind
structures due to co-rotating interaction regions (CIRs)
which are linked to structures
that are rooted at the stellar surface can at least quantitatively
explain the empirical behaviour of the DACs (e.g.
Cranmer {\&} Owocki, 1996; Lobel {\&} Blomme 2008).

Our present investigation of PN fast winds has revealed evidence of DACs in
{\it FUSE} P Cygni line profiles of 5 central stars.
We conclude that the empirical properties 
of these features, including central velocity progression, line-of-sight velocity 
dispersion and recurrence suggest that we are witnessing essentially the same 
physical phenomenon that is commonly evident in massive stars.
Testing relations between DAC time-scales and stellar rotation for CSPNs 
would require considerably more UV datasets, with adequate sampling at multiple 
epochs and for stars with a larger range of rotational velocities.

However, it is instructive to examine the normalised (dimensionless)
DAC
acceleration parameter $a_{DAC} R_\star/v_\infty^2$.  This variable
would be identically constant if DAC absorption originated in CIRs
which
were due to material confined to infinitely thin streak lines (Brown
et
al.\ 2004).  In actuality, the finite extent of the material
responsible
for DACs causes deviations from such a simple scaling law.
Nevertheless, this parameter represents a scaling factor indicative of        
CIR related motion. 
We compared the
values of this parameter derived here for our sample of CSPN
to O star DAC
accelerations obtained from Kaper et al. (1999) data for 68
Cygni, $\xi$
Per, $\lambda$ Cep, $\zeta$ Ori A, 19 Cep, $\lambda$ Ori A, 15 Mon
and 10 Lac. The respective accelerations and adopted parameters
are listed in Table 4.
For the O stars, $0.002 \leq a_{DAC} R_\star/v_\infty^2 \leq 0.059$,
while for the CSPN, we have $0.004 \leq a_{DAC} R_\star/v_\infty^2
\leq
0.071$, where the upper limit is 0.047 if we exclude IC~4593,
for which the DAC progression is poorly constrained
(see Fig. 3.).
We conclude
that the range of this parameter, which is related to CIRs, is
similar for
both CSPNe and OB stars, pointing to a common origin for both.


\begin{table*}
 \centering
\caption{Comparison of normalised DAC accelerations for
CSPN and O stars.}
  \begin{tabular}{lllllll}
  \hline
Star & $R_\star/R_\odot$ & $v_\infty$ (km s$^{-1}$) &
v$\sin i$ (km/s) & a$_{\rm DAC}$ (km s$^{-2}$) 
& $a_{DAC} R_\star/v_\infty^2$ & v$\sin i$/$v_\infty$ \\
\hline

68 Cyg & 16 & 2500 & 300 & 0.0200 & 0.0343 & 0.118 \\
$\xi$ Per & 14 & 2450 & 220 & 0.0170 & 0.0276 & 0.090 \\
$\lambda$ Cep & 21 & 2250 & 200 & 0.0140 & 0.0406 & 0.089 \\
$\zeta$ Ori & 22 & 2100 & 124 & 0.0170 & 0.0590 & 0.059 \\
19 Cep & 23 & 2050 & 100 & 0.0007 & 0.0028 & 0.049 \\
$\lambda$ Ori & 15 & 2400 & 66 & 0.0011 & 0.0020 & 0.028 \\
15 Mon & 9 & 2000 & 67 & 0.0025 & 0.0039 & 0.034 \\
10 Lac & 8 & 1200 & 30 & 0.0018 & 0.0070 & 0.025 \\
 & & & & & & \\
NGC 6826 & 2 & 1200 & 50 & 0.0530 & 0.0457 & 0.042 \\
IC 418 & 2 & 700 & 56 & 0.0160 & 0.0466 & 0.080 \\
IC 4593 & 2 & 950 & 55 & 0.0420 & 0.0706 & 0.058 \\
IC 2149 & 4 & 1000 & 54 & 0.0090 & 0.0242 & 0.054 \\
NGC 6543 & 1 & 1400 & 85 & 0.0220 & 0.0044 & 0.061 \\
\hline
\end{tabular} \\
The O star DAC accelerations (a$_{\rm DAC}$) are
derived from Kaper et al. (1999). The O star parameters
are from Repolust et al. (2004), Markova et al. (2004),
Martins et al. (2005) and Sim{\'o}n-D{\'i}az et al. (2006).
We assign a $\pm$ 25{\%} propagated error for the
normalised accelerations ($a_{DAC} R_\star/v_\infty^2$).
\end{table*}


\section*{Acknowledgments}
D.L.M. acknowledges support from NASA's ADAP Grant NNX11AD28G.
We are grateful for the comments of the referee.

\bsp

\label{lastpage}

\end{document}